\documentclass[aps,prd,twocolumn,nofootinbib,longbibliography,10pt]{revtex4-2}
\usepackage{etoolbox}
\usepackage{dcolumn}
\usepackage{changepage}
\usepackage{amsmath,amssymb,amsfonts,mathtools}
\usepackage{mathrsfs,bbold}
\usepackage{graphicx}
\usepackage[colorlinks=true,urlcolor=blue,citecolor=red,linkcolor=blue]{hyperref}
\usepackage{accents}
\newlength{\dhatheight}
\usepackage{natbib}
\usepackage{wrapfig}
\usepackage{flushend,BOONDOX-cal,BOONDOX-frak}

\makeatletter
\newsavebox{\@brx}
\newcommand{\llangle}[1][]{\savebox{\@brx}{\(\m@th{#1\langle}\)}%
  \mathopen{\copy\@brx\kern-0.5\wd\@brx\usebox{\@brx}}}
\newcommand{\rrangle}[1][]{\savebox{\@brx}{\(\m@th{#1\rangle}\)}%
  \mathclose{\copy\@brx\kern-0.5\wd\@brx\usebox{\@brx}}}
\makeatother

\begin{document}
\title{Detecting gravitational waves by emission of photons from charged Weber bars}
\author{Soham Sen}
\email{sensohomhary@gmail.com}
\affiliation{Institute for Advanced Study, Kyushu University, Fukuoka 819-0395, Japan}
\begin{abstract}
 In this work, we propose a novel experimental set-up using charged resonant gravitational wave detectors. We exploit the semi-classical analogue of the Gertsenshtein effect where the gravitational wave acts as a modulator for the optomechanical system. We consider a cavity QED scenario where the Weber bar is placed inside an electromagnetically shielded cavity. We observe that when the gravitational wave falls on the Weber bar, it emits photon which signifies the detection of gravitational waves by the resonant bars. The frequency controlled photon emission scenario will shed a new light on future generation of efficient gravitational wave detector models. From a detailed strain sensitivity and noise profile analysis, we establish that our proposed detector based on this novel mechanism will be a step forward towards tabletop gravitational wave detection.
\end{abstract}
\maketitle
\section{Introduction}
 Albert Einstein in 1916 proposed the existence of classical gravitational waves \cite{EinsteinGW} which was first detected in 2015 by the LIGO (Laser Interferometer Gravitational-Wave Observatory) gravitational wave detector. The detected gravitational wave signal was generated by the collision of two inspiralling neutron stars \cite{Abott1,Abott2,Abott3}. This first detection and along with simultaneous detection of gravitational waves have led to an upsurge in the research of gravitational wave detection and new gravitational wave detector models. The existing gravitational wave observatories for example LIGO, VIRGO, KAGRA\footnote{VIRGO: Virgo Interferometer for the Detection of Gravitational Waves, KAGRA: Kamioka Gravitational Wave Detector}, and GEO600 are all interferometer based gravitational wave detectors and they can be considered to be one of the most complex engineering marvels. The proposal for the first gravitational wave detector, however, was extremely simple which considers a solid resonant bar and was proposed by James Weber in 1969 \cite{Weber69}. These gravitational wave detectors which relies primarily upon the tiny fluctuation of the quantum matter as a result of small spacetime fluctuations, are famously known as Weber bar or resonant bar detectors. There have been a series of works investigating the quantum mechanical response of a resonant bar towards incoming gravitation waves from astrophysical sourced\cite{RD1,RD2,RD3,RD4,RD5,RD6}.

 Very recently, in \cite{Schutzhold}, a proposal for an optical Weber bar has been given where the exchange of energies between the gravitational as well as electromagnetic waves was investigated in an extended Mach-Zehnder or Sagnac type geometry. This is a very interesting effect present in nature which is also known as the Gertsenshtein effect \cite{Gertsenshtein} where a gravitational wave and electromagnetic wave interchange into each other in the presence of a background magnetic field. In a semi-classical scenario where the electromagnetic field is showing strong quantum nature, however, one should observe a semi-classical version of the classical Gertsenshtein effect where the energy of the gravitational wave will convert into photons with a suitable resonance condition. In our set-up, however, we are proposing a more intricate model where a charged or conducting resonant bar is placed inside a photon-shielded cavity with the cavity filled with electromagnetic radiation (for a more realistic set-up one can also consider a background magnetic field in presence). If the gravitational wave enters this optomechanical set-up, it will create tiny vibrations in the elastic quantum matter where the vibrations behave as quasi-particle states, also called phonons. The aim is to look for if the energy transferred from the gravitational wave creates an excited phonon with a simultaneous emission of a photon and if electromagnetic pumping can enhance this signal experimentally. The emission of photons while the perfect resonance condition is satisfied will lead towards the direct detection of gravitational waves by Weber bars. The primary benefit of this process will be the verification of the semi-classical Gertsenshtein effect as well as the controlled detection scenario of gravitational waves in a cavity optomechanical set-up which is both more efficient and easy to implement. The paper is organized as follows.

 In sec.(\ref{Action}), we discuss the physical model and derive the action and eventually the Hamiltonian for the matter system. In sec.(\ref{QuantumGertsenshtein}), we discuss in the details the photon emission scenario and the stimulated as well as spontaneous emission case for the photons. In subsection (\ref{ExperimentalProposal}), we propose a simple optomechanical set-up for detecting gravitational fluctuations and finally in sec.(\ref{Conclusion}), we summarise our results.
\section{Action for the model system}\label{Action}
 We model the resonant bar by considering the collective mass of the vibrating particles as a single particle with mass $m_0$ while it is connected to a heavier mass particle $m_\infty$ while connected by a massless spring with oscillation frequency $\omega_0$. The spring length $\xi=\sqrt{\xi_i\xi^i}$ is effectively the geodesic separation between the two masses $m_0$ and $m_\infty$. If one now considers that the heavier mass is following a time-like geodesic then the coordinates of the particle with smaller mass is given simply by the Fermi-normal coordinates$\mathcal{Y}^\mu=\{t,\xi^i\}$ \cite{Fermi_Normal,Fermi_Normal2}. This coordinate is extensively used also in simple gravitational wave as well as graviton detector models \cite{QGravD,QGravLett,KannoSodaTokuda,
 KannoSodaTokuda2,OTMGraviton}. 
The total action for the system will comprise of the action for the two particles $m_0$ as well as $m_{\infty}$. Now, as $m_\infty$ follows a time-like geodesic and is placed as the origin of the Fermi-normal coordinates the dynamics for $m_\infty$ can simply be ignored. One can approximately write down the action for the model system as
\begin{equation}\label{II.2}
\begin{split}
S_{\text{RD}}^{(0)}\simeq-m_0\int dt \left(\sqrt{-g_{\mu\nu}\frac{d\mathcal{Y}^\mu}{dt}\frac{d\mathcal{Y}^\nu}{dt}}+\frac{1}{2}\omega_0^2 g_{\mu\nu}\mathcal{Y}^\mu\mathcal{Y}^\nu\right)
\end{split}
\end{equation}
with the oscillation frequency of system being $\omega_0$. 
\begin{figure}
\begin{center}
\includegraphics[scale=0.9]{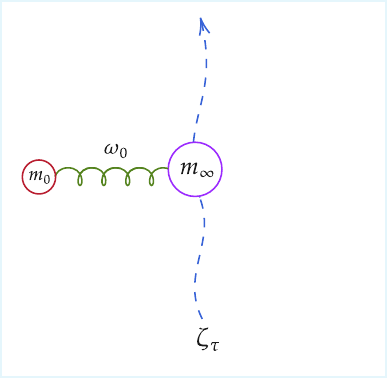}
\caption{The resonant bar is modelled by a smaller mass particle of mass $m_0$ connected to a heavier mass particle with mass $m_\infty$ where the two particles are connected by a spring with oscillation frequency $\omega_0$ and the particle with mass $m_\infty$ follows a time-like geodesic $\zeta_\tau$. \label{Bar_Model_OTM}}
\end{center}
\end{figure}
For a conducting detector carrying a finite charge $q$, we can consider that the charge is carried entirely by the mass $m_0$ for simplicity of analysis and it will couple to the the electro-magnetic field $A_\mu$ via the standard minimal coupling between the charged particle and electromagnetic field and the corresponding action reads
\begin{equation}\label{II.3}
\begin{split}
S_{\text{RD}}^q=&q\int A_{\mu}d\mathcal{Y}^{\mu}
=q\int dt~ g_{\mu\nu}A^\mu \dot{\mathcal{Y}}^\nu~.
\end{split}
\end{equation}
Setting the Coulomb gauge condition ($A^0=0$ and $\vec{\nabla}\cdot\vec{A}=0$), we can write down the full analytical action for the resonant bar in the Fermi-normal coordinates as 
\begin{equation}\label{II.4}
\begin{split}
S_{\text{RD}}
=&\frac{m_0}{2}\int dt \left(\delta_{ij}\dot{\xi}^i\dot{\xi}^j-R_{i0j0}(t,0)\xi^i\xi^j-\omega_0^2\delta_{ij}\xi^{i}\xi^j\right)\\
-&\frac{2}{3}q\int dt~ R_{0jik}(t,0)\xi^j\xi^k A^i+q\int dt~\delta_{jk}A^j\dot{\xi}^k
\end{split}
\end{equation}
where the Riemann curvature tensor reads $R_{i0j0}(t,0)\simeq -\frac{1}{2}\ddot{\bar{h}}_{ij}(t,0)$ and $R_{0jik}(t,0)\simeq \frac{1}{2}\left(\partial_j\dot{\bar{h}}_{ik}-\partial_i\dot{\bar{h}}_{jl}\right)$ up to $\mathcal{O}(h)$. The action corresponding to the electromagnetic field is given by the Maxwell action and its curved spacetime analogue in the Coulomb gauge reads
\begin{equation}\label{II.5}
\begin{split}
S_{\text{M}}=&-\frac{1}{4} \int d^4x \sqrt{-g} g^{\mu\alpha}g^{\nu\beta} F_{\alpha\beta} F_{\mu\nu}\\
\simeq &\frac{1}{2}\int d^4x \left(\eta^{ij}\dot{A}_i\dot{A}_j-\eta^{ij}\eta^{kl}\partial_iA_k\partial_j A_l\right)+S_{\text{M}}^h
\end{split}
\end{equation}
where $F_{\mu\nu}=\nabla_{\mu}A_\nu-\nabla_{\nu}A_\mu=\partial_\mu A_\nu-\partial_\nu A_\mu$  and $S_{\text{M}}^h$ is given by
\begin{equation}\label{II.6}
S_{\text{M}}^h=-\frac{1}{2}\int d^4x~ \bar{h}^{ij}(t,\vec{x})\left(\dot{A}_j\dot{A}_j-\eta^{kl}F_{ki}F_{lj}\right)~.
\end{equation}
Our primary aim here is to look for the interaction of the classical gravitational wave with the bar-EM field coupled system where the degrees of freedom corresponding to the Weber bar and the electromagnetic field is treated quantum mechanically.
\subsection{Quantizing the detector as well as the electromagnetic part}  
 We start by the discrete mode decomposition for the electromagnetic field vector $A_i(t,\vec{x})$ as (where in the Coulomb gauge $A_0=0$) 
\begin{equation}\label{II.7}
A_{i}(t,\vec{x})=\frac{1}{\sqrt{\hbar G^2}}\sum_{\vec{k}_P,P}A_{P}(t,\vec{k}_P)e^{i\vec{k}_P\cdot\vec{x}}\epsilon^s_{i}(\vec{k}_P)
\end{equation}
with $A_{P}(t,\vec{k}_P)$ being the Fourier mode function and $\epsilon^s_i(\vec{k}_P)$ denoting the electromagnetic polarization tensor. We now consider that the transverse wave vector of the gravitational wave is parallel to the electromagnetic wave, that is $\vec{k}\parallel \vec{k}_P$ and the bar is aligned perpendicular to the direction of the propagation of the gravitational as well as electromagnetic wave. If $\vec{k}=\{0,0,k\}$ and $\vec{k}_P=\{0,0,k_P\}$ then we consider the geodesic separation to be $\xi^i=\{\xi,\delta\xi^y,\delta\xi^z\}$ as the height and the width of the resonant bar detector is negligible with respect to its length. It is therefore evident that $\xi\gg \delta\xi^y,\delta\xi^z$. We can therefore restrict our model primarily to the $x$ direction and neglect any higher order as well as dynamical contributions from the perpendicular directions $\mathcal{O}(\delta\xi^2,\delta\dot{\xi})$. In the long-wavelength approximation and restricting to a single mode of the electromagnetic wave, the Lagrangian for the model system can be read off from the total action ($S=S_{\text{RD}}+S_{\text{M}}$) as
\begin{widetext}
\begin{equation}\label{II.8}
\begin{split}
L=&\frac{m_0}{2}\left(\dot{\xi}^2-\dot{\bar{h}}_{xx}(t,0)\dot{\xi}\xi-\omega_0^2\xi^2\right)+q_PA\dot{\xi}-\frac{\delta\xi^z q_P}{3}\partial_z\dot{\bar{h}}_{xx}(t,z)\rvert_{z\rightarrow 0}\xi A+\frac{m_P}{2}(1-\bar{h}^{xx}(t,0))(\dot{A}^2-\omega_P^2A^2)
\end{split}
\end{equation}
\end{widetext}
where $m_P=\frac{L^3}{\hbar G^2}$, $\Re[A_+(t,\vec{k}_P)]=A$, $\Im[A_+(t,\vec{k}_P)]=0$ while considering only the plus polarization for the electromagnetic wave mode and $q_P\equiv \frac{q}{\sqrt{\hbar G^2}}$.
We can now consider a plane polarized gravitational wave with a single mode $k=\omega$ ($c\rightarrow1$) propagating in the $z$ direction and this helps us to write $\bar{h}_{ij}(t,\vec{x})$ as
\begin{equation}\label{II.9}
\begin{split}
\bar{h}_{ij}(t,\vec{x})=&\sum_s h_s(t,\vec{k})\epsilon^s_{ij}(\vec{k})\cos(\omega t-\vec{k}\cdot\vec{x})\\=&2f_0 \epsilon^+_{ij}(k)\cos(\omega t-k z)~.
\end{split}
\end{equation}
We are now in a position to construct the Hamiltonian from the Lagrangian by analytically obtaining the conjugate to $\xi$ and $A$ as $\pi_\xi=\frac{\partial L}{\partial \dot{\xi}}$ and $p_A=\frac{\partial L}{\partial \dot{A}}$, raising all the phase space variables to operator status and implementing suitable canonical commutation relations between the conjugate pairs, that is $\{\hat{\xi},\hat{\pi}_\xi\}$ and $\{\hat{A},\hat{p}_A\}$. We can now write down the Hamiltonian operator for the entire model system up to $\mathcal{O}(h)$ as
\begin{equation}\label{II.10}
\begin{split}
\hat{H}=&\mathbb{1}_{\text{P}}\otimes\hat{H}^0_{\text{RD}} +\hat{H}^0_{\text{P}}\otimes \mathbb{1}_{\text{RD}}+\frac{1}{2}\omega f_0 \sin\omega t(\hat{\pi}_\xi\hat{\xi}+\hat{\xi}\hat{\pi}_\xi)\\-&\frac{q_P}{m_0}\hat{A}\otimes \hat{\pi}_\xi+2f_0\cos\omega t\left(\frac{\hat{p}_A^2}{2m_P}-\frac{1}{2}m_P\omega_P^2\hat{A}^2\right)\\
-&q_P\omega f_0 \hat{A}\otimes \hat{\xi}\left(\sin\omega t-\frac{2\omega\delta\xi^z}{3}\cos\omega t\right)
\end{split}
\end{equation}
where $\hat{H}^0_{\text{RD}}=\frac{\hat{\pi}_\xi^2}{2m_0}+\frac{1}{2}m_0\omega_0^2\hat{\xi}^2$, $\hat{H}^0_{\text{P}}=\frac{\hat{p}_A^2}{2m_P}+\frac{1}{2}m_P\Omega_P^2\hat{A}^2$, and $\Omega_P^2\equiv \omega_P^2+\frac{q_P^2}{m_0m_P}$\footnote{Here, $\Omega_P\simeq \omega_P$ if and only if  $\omega_P\gg \frac{q_P}{\sqrt{m_0m_P}}$.}. 

 With the analytical form of the Hamiltonian in eq.(\ref{II.10}), we are now in a position to investigate the physical interpretation of the above model system. The first interaction term $\frac{1}{2}\omega f_0 \sin\omega t(\hat{\pi}_\xi\hat{\xi}+\hat{\xi}\hat{\pi}_\xi)$ is the well known gravitational wave- Weber detector interaction term. This term actually is the term that will allow the detector to excite and jump two energy levels by absorbing energy from the gravitational wave provided the resonance condition $\omega=2\omega_0$ is satisfied. The next interaction term $-\frac{q_P}{m_0}\hat{A}\otimes \hat{\pi}_\xi$ signifies the photon-phonon conversion term introducing the excitation and de-excitation of the detector via absorbing and emitting photons. Finally, the term $2f_0\cos\omega t\left(\frac{\hat{p}_A^2}{2m_P}-\frac{1}{2}m_P\omega_P^2\hat{A}^2\right)$ leads to the Gertsenshtein effect, however, as the electromagnetic waves are quantized, the gravitational wave will convert into two photons as a result of this interaction term where the resonant detector remains unchanged by this interaction provided the $\omega=2\Omega_P$ resonance condition gets satisfied. In \cite{Schutzhold}, this term is primarily investigated to propose the ``optical Weber bar" model. In our analysis, however, we are most interested in the final interaction term in eq.(\ref{II.10}). We can consider $\omega\delta\xi^z\ll 1$ and write down the three party interaction term simply as
$\hat{\mathcal{H}}_{\text{int}}\simeq -q_P\omega f_0 \sin\omega t\hat{A}\otimes \hat{\xi}$. The interesting point to understand is that this interaction term indeed involves the detector, photons as well as the gravitational wave.
\section{The three mode interaction and its novel physical aspects}\label{QuantumGertsenshtein}
\subsection{Stimulated and spontaneous emission of photons from the charged detector}
 With the physical model in place, we can now obtain the transition amplitude and transition probability of the model system for going from an initial state $|\psi_i\rangle$ to some final state $|\psi_f\rangle$. We start our analysis by considering that the detector is in its ground state and $n_{P_i}$ number of photons are there initially in the system.
The initial state of the system then reads $|\psi_i\rangle=|n_{P_i},0\rangle$. We shall now look at the feasible transitions allowed by the interaction Hamiltonian $\hat{\mathcal{H}}_{\text{int}}$. At first, we need to write the interaction Hamiltonian $\hat{\mathcal{H}}_{\text{int}}$ in the interaction picture, which is obtained by writing all the phase space operators in the interaction picture and they are given by
$\hat{\xi}^I=\sqrt{\frac{\hbar}{2m\omega_0}}\left(\hat{\chi}e^{-i\omega t}+\hat{\chi}^\dagger e^{i\omega t}\right)$, $\hat{\pi}_\xi=i\sqrt{\frac{m_0\hbar\omega_0}{2}}(\hat{\chi}^\dagger e^{i\omega_0 t}-\hat{\chi}e^{-i\omega_0 t})$, 
$\hat{A}^I=\sqrt{\frac{\hbar}{2m_P\bar{\omega}_P}}\left(\hat{a}e^{-i\Omega_P t}+\hat{a}^\dagger e^{i\Omega_Pt}\right)$, and $\hat{p}_A=i\sqrt{\frac{m_P\hbar \Omega_P}{2}}\left(\hat{a}^\dagger e^{i\Omega_P t}-\hat{a} e^{-i\Omega_P t}\right)$ with $\hat{\chi}$ and $\hat{\chi}^\dagger$ denoting the lowering a raising operators corresponding to the Weber bars, and $\hat{a}$ and $\hat{a}^\dagger$ the annihilation and creation operators for the photon with frequency $\Omega_P$.
The interaction Hamiltonian $\hat{\mathcal{H}}_{\text{int}}$ in the interaction picture then reads
\begin{equation}\label{II.11}
\begin{split}
&\hat{\mathcal{H}}_{\text{int}}^I=-\frac{\hbar \omega q_P f_0}{4i\sqrt{m_P\Omega_Pm_0\omega_0}}(e^{i\omega t}-e^{-i\omega t})\left(\hat{a}\hat{\chi}e^{-i(\omega_0+\Omega_P)t}\right.\\&\left.+\hat{a}\hat{\chi}^\dagger e^{-i(\omega_0-\Omega_P)t}+\hat{a}^\dagger\hat{\chi} e^{i(\omega_0-\Omega_P)t}+\hat{a}^\dagger\hat{\chi}^\dagger e^{i(\omega_0+\Omega_P)t}\right)~.
\end{split}
\end{equation}
The transition amplitude for the above interaction Hamiltonian up to first order in the interaction Hamiltonian simply reads
\begin{equation}\label{II.12}
\langle \psi_f |\hat{\mathcal{U}}^I(t_f,t_i)|\psi_i\rangle\simeq-\frac{i}{\hbar}\int_{t_i}^{t_f} dt' \langle \psi_f|\hat{\mathcal{H}}_{\text{int}}^I|\psi_i\rangle~.
\end{equation}
For better analytical understanding, we consider the $t_i\rightarrow -\infty$ and $t_f\rightarrow\infty$ limit. Here, several physical phenomena can occur depending on the realization of different resonance conditions.  For $\omega=\omega_0+\Omega_P$, only two terms contribute rendering the transition probability $\mathcal{P}_{if}=\left|\langle \psi_f |\hat{\mathcal{U}}^I(t_f,t_i)|\psi_i\rangle\right|^2$ to have the form
\begin{widetext}
\begin{equation}\label{II.13}
\begin{split}
\mathcal{P}_{if}=&\frac{\pi^2\omega^2 q_P^2f_0^2}{4m_P\Omega_Pm_0\omega_0}\left|\langle \psi_f |\hat{a}^\dagger \hat{\chi}^\dagger|\psi_i\rangle-\langle \psi_f |\hat{a} \hat{\chi}|\psi_i\rangle\right|^2\delta^2(\omega-\omega_0-\Omega_P)\\
=&\frac{\pi^2\omega^2 q_P^2f_0^2}{4m_P\Omega_Pm_0\omega_0}(n_{P_i}+1)\delta_{n_{P_f},n_{P_i}+1}\delta_{n_{R_f},1}\delta^2(\omega-\omega_0-\Omega_P)
\end{split}
\end{equation}
\end{widetext}
where the Dirac delta function ensures that of the resonance condition does not get satisfied the entire transition probability goes away whereas the Kronecker deltas ensure that the transition probability is non zero provided the system has jumped to its first excited state while simultaneously emitting a photon with the final state of the system being $|\psi_f\rangle=|n_{P_f},n_{R_f}\rangle=|n_{P_i}+1,1\rangle$. 
 The final analytical form of the transition probability then reads
$\mathcal{P}_{if}=\frac{\pi^2\omega^2 q_P^2f_0^2}{4m_P\Omega_Pm_0\omega_0}(n_{P_i}+1)\delta^2(\omega-\omega_0-\Omega_P)$\footnote{It is important to note that all the unphysical processes containing the term $\delta^2(\omega+\omega_0+\Omega_P)$ has been dropped throughout the entire analysis.}. 
\subsection{Parameter Estimation}
In this subsection, we shall proceed with parameter estimation for a valid experimental proposal. The dimensionally restored form of the transition probability while considering the entire model in SI units read
\begin{equation}\label{II.14}
\mathcal{P}_{if}=\frac{\pi^2\omega^2 q^2f_0^2}{4\epsilon_0L^3\Omega_Pm_0\omega_0}(n_{P_i}+1)\delta^2(\omega-\omega_0-\Omega_P)
\end{equation}
with $\epsilon_0=8.854\times 10^{-12}$ $\text{F}.\text{m}^{-1}$ being the permittivity of free space. One important thing to remember is that the transition rate is an experimentally observable quantity and in an experimental scenario the observation is executed for a finite time say $\tau$ then the delta function can be replaced by $2\pi\delta(\omega-\omega_0-\Omega_P)\rightarrow \int_{-\frac{\tau}{2}}^{\frac{\tau}{2}}dt e^{i(\omega-\omega_0-\Omega_P)t}=\frac{2}{(\omega-\omega_0-\Omega_P)}\sin\left[(\omega-\omega_0-\Omega_P)\frac{\tau}{2}\right]$ which at resonance point gives $\delta(\omega-\omega_0-\Omega_P)\rightarrow \frac{\tau}{2\pi}$. The transition rate then simply becomes $\Gamma_{if}=\frac{1}{\tau}\mathcal{P}_{if}=\frac{\pi\omega^2 q^2f_0^2}{8\epsilon_0L^3\Omega_Pm_0\omega_0}(n_{P_i}+1)\delta(\omega-\omega_0-\Omega_P)$ which at the resonance point exactly reads $\frac{\omega^2 q^2f_0^2 \tau}{16\epsilon_0L^3\Omega_Pm_0\omega_0}(n_{P_i}+1).$

We consider a resonant detector of length $L=10^{-2}$ m and mass of $m_0=10^{-6}$ kg (Detector made of Aluminium), which sets the effective fundamental frequency to $\omega_0=2\pi\nu=\frac{v_{\text{Sound}}}{2L}=\frac{3000}{2\times10^{-2}} \text{ Hz}=1.5 \times 10^5$ Hz. For a resonant bar the collective phonon mode frequency actually result in the fundamental frequency of the bar and as a result, the frequency is calculated using the speed of sound which is $3000-5000$ $\text{m.sec}^{-1}$ depending on the purity of the material. For reference, if we consider an incoming gravitational wave with frequency $\omega=2\times 10^5$ Hz then the emitted photon frequency ($\Omega_P$) due to the resonance condition reads $\Omega_P=\omega-\omega_0=5\times 10^4$ Hz. For an incoming gravitational wave the dimensionless amplitude lies in the range $f_0\sim 10^{-21}$. We consider oscillators placed inside of electromagnetically shielded cavities. On one milligram of mass, one can put a maximum charge $q\sim 10^{-4}-10^{-5}$ C and we proceed with an experimentally achievable  $q=10^{-4}$ C charge. The total observation time for our experimental proposal is set to approximately $\tau\sim 10^5$ sec. If we now consider the spontaneous emission case for photons then, one needs to set $n_{P_i}=0$. This sets the spontaneous emission rate for photons to $\Gamma_{if}\simeq 3.76\times 10^{-23}$ $\text{sec}^{-1}$.  Hence, if a charged detector absorbs a gravitational wave higher than its fundamental oscillation frequency it jumps to its first excited state while emitting a single photon spontaneously and this spontaneous emission rate is given by $\Gamma_{if}\simeq 3.76 \times10^{-23}$ $\text{sec}^{-1}$ which is extremely low.
Thus the observation of spontaneous emission of photons purely due to the semi-classical Gertsenshtein effect is not experimentally feasible while considering a single oscillator with effective mass $m_0=10^{-6}$ kg. 
There are now few distinct ways to boost the transition rate significantly. The first way is to introduce optical or electromagnetic pumping and second way is to use squeezed phonon modes. 

A completely reasonable physical scenario can be observed when the cavity is electromagnetically pumped hugely increasing the number of photons initially present inside of the cavity before the interaction of the detector starts with the gravitational wave. In such a scenario one can effectively increase $n_{P_i}$ up to $10^{19}-10^{22}$ which bumps the transition rate for stimulated emission to $\Gamma_{if}\sim 10^{-4}-0.1$ $\text{sec}^{-1}$ which hugely boosts detectability of the gravitational waves using charged resonant oscillators inside of a high Q cavity via realization of the semi-classical Gertsenshtein effect. Such a large value for the optical pumping will not lead to significant radiation pressure and the temperature gain is also minimal. If we consider $10^{22}$ photons with frequency $\omega_P=5\times 10^4$ Hz (which is required by the resonance condition), the total change in energy due to optical pumping reads $E_{\text{Pumping}}=n_{P_i}\hbar\Omega_P=\frac{1}{2\pi}10^{22}\times 6.626\times 10^{-34}\times 5\times10^4 \text{J}\simeq 5.27\times 10^{-8}$ J. Even for a box of volume $V_{\text{System}}=1$ $\text{cm}^3$, the energy density becomes $u_{\text{Rad.}}=\frac{E_{\text{Pumping}}}{V_{\text{System}}}=\frac{5.27\times 10^{-8}}{10^{-6}}=5.27\times 10^{-2}$ Pa. The radiation pressure is proportional to energy density. For a perfectly absorbing surface $P_{\text{Rad.}}=u_{\text{Rad.}}$ and for a perfectly reflecting surface the radiation pressure is $P_{\text{Rad.}}=2u_{\text{Rad.}}$. For our case, it is more prudent to proceed with an isotropic radiation case where $P_{\text{Rad.}}=\frac{u_{\text{Rad.}}}{3}$ which brings the radiation pressure for optical pumping to $P_{\text{Rad.}}\simeq 1.76 \times 10^{-2}$ Pa which is way smaller than the atmospheric pressure and can be easily controlled in an advanced experimental scenario ($\frac{P_{\text{Rad.}}}{P_{\text{Atmos.}}}\sim10^{-8}$). For a system volume $V_{\text{System}}=1$ $\text{m}^3$ the radiation pressure further reduces to $P_{\text{Rad.}}\simeq 1.76 \times 10^{-8}$ which is extremely small and does not lead to any intricacies during experimental implementation. For an experimental implementation of electromagnetic shielding, we need an advanced Faraday cage, which will be made of materials like copper or aluminium. For consideration, copper has a specific heat of $C_{\text{Cu}}=385$ $\text{J}\cdot\text{kg}^{-1}\cdot\text{K}^{-1}$. Now considering the mass of the cavity wall set-up to be $M=1$ Kg, the temperature gain by the cavity is simply (from the first law of thermodynamics) $\Delta U=U_{\text{Rad.}}=M C_{\text{Cu}}\Delta T_{\text{Rad.}}\implies \Delta T=\frac{U_{\text{Rad.}}}{M C_{\text{Cu}}}\simeq 1.37\times 10^{-10} $ K=137 pK which is extremely small. If the set-up weight is larger, then the temperature gain becomes significantly smaller. Such minute change of temperature can be easily controlled by putting the cavity-QED set-up inside of a cryogenic chamber \cite{CryogenicChamber}. If one now introduces squeezed phonon modes with a real squeezing parameter $r$, an effective gain of $e^{2r}$ in the transition rate as well as transition probability is observed for $r>1$. Still now a maximum of $d=11.5$ dB squeezing is reported which is equivalent to a real squeezing $d=-10\log_{10}[e^{-2r}]\implies r=\frac{d}{20}\ln[10]\sim 1.32$ \cite{SqueezingPRL}. Recent proposal for optical squeezing up to 28 dB \cite{PhotonSqueezing} has also been put forward which is equivalent to a $r=3.22$ squeezing. If such squeezing can be achieved for mechanical oscillators a direct gain of the order of $10^2$ can be achieved in the transition rate.
\subsection{Experimental Implementation}
\label{ExperimentalProposal}

\begin{figure}
\begin{center}
\includegraphics[scale=0.7]{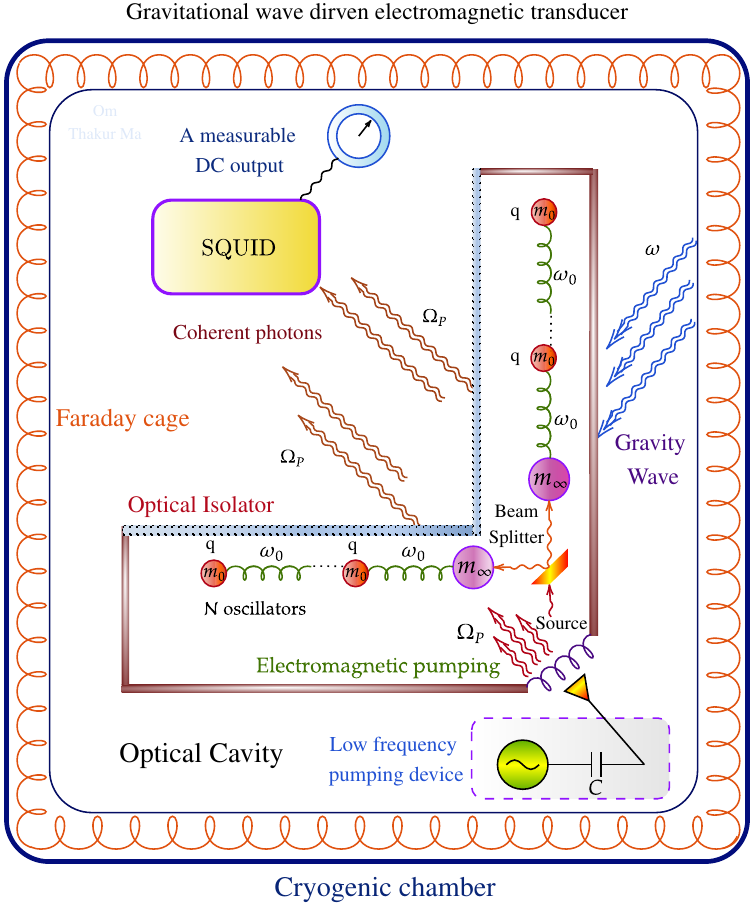}
\caption{A schematic diagram (not to scale) of an array resonant bar detector based on the principle of semiclassical Gertsenshtein effect where the mechanical coherent array of oscillators act as a transducer converting the gravitational wave into low frequency photons while simultaneously getting excited. This collective signal is then converted into DC output signal using a superconducting quantum interference device.\label{SQUID_GW_OTM}}
\end{center}
\end{figure}

 Consider a high Q cavity inside which an array of small oscillators are placed. Such a scenario can be observed using spring oscillator connected in array which results in a resultant amplification of the order of $\mathcal{N}^2$ to the transition rate where $\mathcal{N}$ denotes the number of oscillators connected in an array. Such a model can be implement using modern parallel crystal resonators. Instead of the detection of spontaneous emission of photons, one can transform the collective weak electromagnetic signal into a measurable current. 
This indeed allows for a indirect detection of gravitational waves using a controlled experimental set up. Enhancement to the overall signal can be done using arrays with identical emission output coupled using beam splitters. We have given a schematic diagram of the experimental proposal in this work in Fig.(\ref{SQUID_GW_OTM}). The first step is to create an array of charged resonant bars which results in an overall amplification of the transition rate. Multiple such arrays can be made fully coherent using a multi-beam splitter and using the separated beams to fall on such identical arrays. The next step is to construct a low-frequency electromagnetic pumping device. 
The easiest way is to create the cavity using resonant oscillators such that the frequency of the cavity walls become identical to the photon frequency $\Omega_P$ which allows for substantial resonant emission of photons inside of the cavity. The final step is to use superconducting quantum interference device or SQUID \cite{SQUIDGW} to capture the emitted photon signal and convert them into a measurable DC output signal resulting in an indirect detection of gravitational waves. If $\mathcal{N}=100$ oscillator array is created, an overall gain of the order of $10^4$ can be achieved in a realistic experimental scenario. Implementing identical $n$ arrays will lead to an effective gain of $n\mathcal{N}^2$ in the transition rate. This experimental set up allows for detection of the semiclassical three point Gertsenshtein effect with an effective transition rate $\Gamma_{if}\sim 0.1-10^{2}$ $\text{sec}^{-1}$ depending on an optical pumping of $n_{P_i}\sim 10^{17}-10^{20}$ for a fully coherent 10 array system with each array consisting of 100 identical oscillators. This is a highly detectable scenario and will pave a new tabletop experiments for gravitational wave detection. The transition rate can be enhanced further via implementing squeezing in the phonon modes but that remains to be an experimentally implementable challenge in our currently proposed model detector.
One can also propose suitable experiments based on the other resonance conditions $\omega_0=\omega+\Omega_P$ and $\Omega_P=\omega+\omega_0$, however, to actually detect gravitational waves using the above two scenarios are more experimentally challenging.
\section{Strain sensitivity analysis}
The strain sensitivity analysis will tell us whether it is possible to detect true gravitational waves using our gravitational wave detector model.
The expected number of events is related to the transition rate for a total observation time $\tau_{\text{Obs}}$, simply by the relation $N_{\text{Event}}=\Gamma_{\text{Emission}} \tau$. For the minimum sensitivity, we consider number of events to be $N_{\text{Event}}=1$, which lets us write down the expression for strain amplitude (for the spontaneous emission case) to be
\begin{equation}\label{S.1.OTM}
\mathcal{f}_{\text{min}}=\sqrt{\frac{16\epsilon_0 L^3\Omega_Pm_0\omega_0}{\omega^2 q^2 \tau^2}}~.
\end{equation} 
For the used parameter values in the previous section, the minimum characteristic strain sensitivity of the detector is obtained to be $\mathcal{h}_{\text{min}}=\frac{\mathcal{f}_{\text{min}}}{\sqrt{\omega}}\simeq 1.15\times 10^{-15}$ $\text{Hz}^{-\frac{1}{2}}$ which indicates that using the current model spontaneous emission of photons due to gravitational wave interaction with the detector is impossible. If we now instead consider the optical pumping scenario, the strain amplitude becomes $\mathcal{f}_{\text{min}}=\sqrt{\frac{16\epsilon_0 L^3\Omega_Pm_0\omega_0}{(n_{P_i}+1)\omega^2 q^2 \tau^2}}$ which for $n_{P_i}\sim 10^{18}$ gives a minimum value of the strain sensitivity to be $\mathcal{h}_{\text{min}}=\frac{\mathcal{f}_{\text{min}}}{\sqrt{\omega}}\simeq 1.15\times10^{-24}$ $\text{Hz}^{-\frac{1}{2}}$ which is a very reasonable range for a gravitational wave detector and is of the order of the existing gravitational wave observatories like LIGO or VIRGO. For $n_{P_i}=10^{20}$, the strain sensitivity becomes $\mathcal{h}_{\text{min}}\simeq 1.15\times10^{-25}$  $\text{Hz}^{-\frac{1}{2}}$ which is better than the existing gravitational wave observatories.  With an $n$ array detector set-up with $\mathcal{N}$ oscillators each, the strain sensitivity becomes more significant reducing the dependence on the optical pumping.
\subsection{Noise profile analysis}
We now need to carefully consider few of the noise sources that needs to be carefully considered for our current experimental proposal. We are considering the detector with a fundamental frequency $\omega_0=1.5 \times 10^5$ Hz and the emitted photon frequency is $\Omega_P=5\times10^4$ Hz. The number of thermal photons at a temperature $T=4K$ is simply $n^{\text{Th.}}_P=\frac{1}{e^{\frac{\hbar\omega_P}{k_BT}}-1}\simeq 1.05\times 10^7$. This number can be significantly reduced by considering an ultra cold or cryogenic cavity inside which the detector is built. For a $1 \mu\text{K}$ set-up $n_{\text{Th.}}\simeq 2$ which is significantly smaller resulting in the deduction of the noise due to thermal emission. We are considering a high-Q cavity and a maximum mechanical quality factor of the order of $10^7$ was obtained in \cite{MechanicalQ} almost two and a half decades back. The mechanical damping rate then simply reads $\Gamma_{\text{Damping}}=\frac{\omega_0}{2Q}\sim 10^{-2}$ Hz. For state of the art oscillators (cryogenic oscillators) it may be possible to attain a $Q$ factor in the range $10^9$ which will reduce the damping factor further by two orders of magnitude ($\Gamma_{\text{Damping}}\sim 10^{-4}$ Hz). The other case is the dark count rate or the superconducting detector suddenly produces a false count which for state of the art SQUID detectors is of the order of $\Gamma_{\text{DCR}}\sim 10^{-4}-10^{-5}$ Hz. As a result the effective signal to noise ratio can be obtained as $\text{SNR}\sim\frac{\Gamma_{if}}{\Gamma_{\text{Damping}}+\Gamma_{\text{DCR}}+\Gamma_{\text{Th.}}}\sim \frac{0.1}{10^{-2}}\sim 10 $ for current mechanical oscillators with $Q$ factor $10^7$, however, for state of the art detectors and with an emission rate of 0.1 Hz the $\text{SNR}~10^3$ which provides a solid ground for the experimental proposal in our work. We now need to consider the noise profile for the SQUID detector.
\subsection{SQUID analysis}
\begin{figure}
\begin{center}
\includegraphics[scale=0.3]{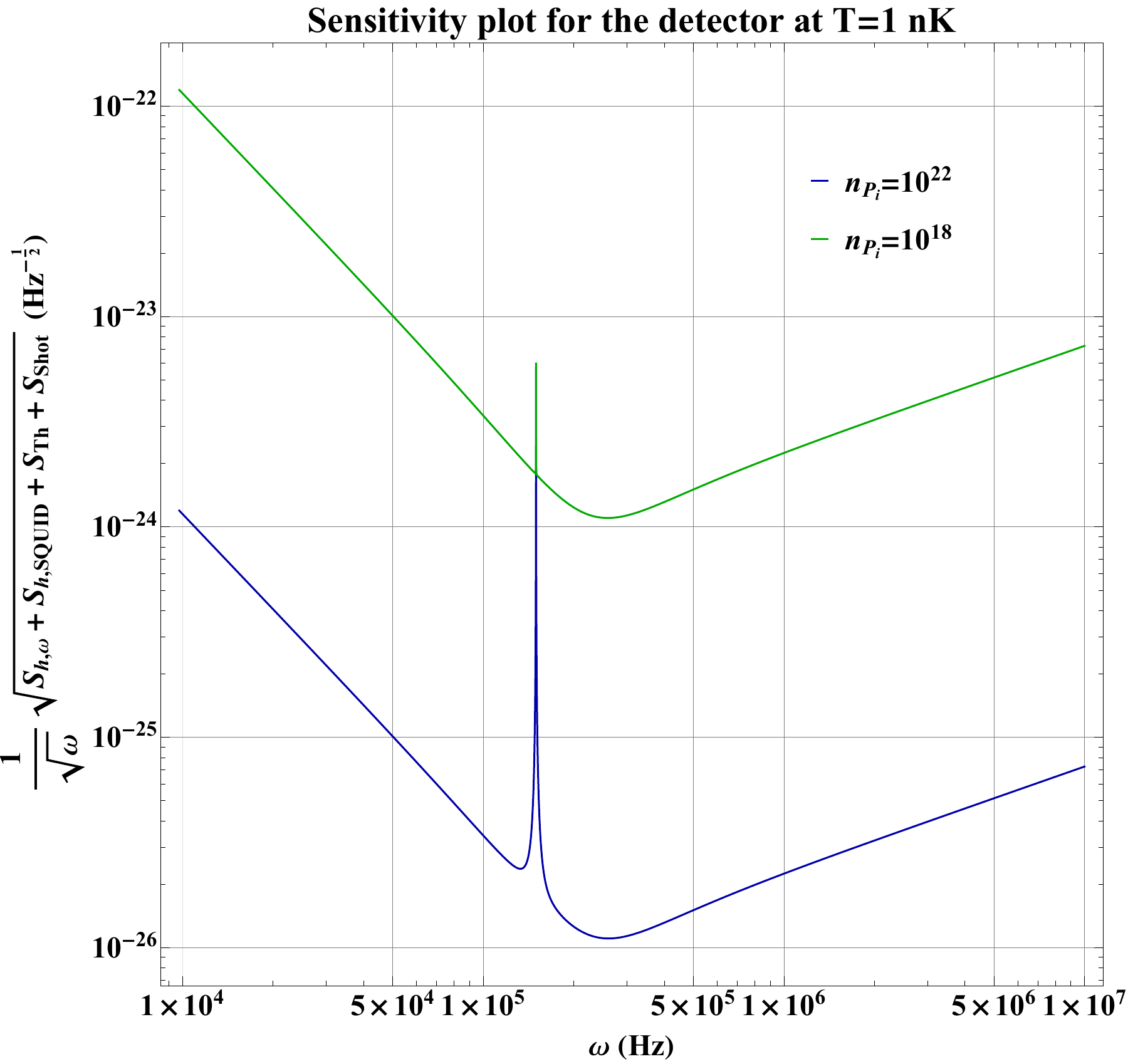}
\caption{The sensitivity plot for our proposed experimental model against the frequency $\omega$. \label{Strain_Sensitivity_OTM}}
\end{center}
\end{figure}

The flux noise for any SQUID detector universally follows the relation
\begin{equation}\label{SQUID.1.OTM}
\Phi_{\text{RMS,Noise}}=\mu\Phi_0 \frac{\sqrt{\Delta \omega}}{\sqrt{Hz}}
\end{equation}
 where $\Phi_0=h/2e$ denotes the fundamental magnetic flux quantum and $\sqrt{\Delta\omega}$ denotes the bandwidth. The bandwidth $\Delta\omega$ is proportional to the inverse of the observation time $\tau$ and for the highest resolution $\Delta\omega_{\text{min}}=\frac{1}{\tau}$. For a good SQUID device $\mu=10^6-10^7$, however, very recently a very low flux noise SQUID has been designed with $\mu=50\times 10^{-9}$\cite{NatureSQUID}. For a rigorous dimensional check we define a parameter $\nu_F=1 Hz$ and this lets us write the RMS noise flux as $\Phi_{\text{RMS,Noise}}=\frac{1}{\sqrt{\tau\nu_F}}\mu\Phi_0 $.
The number of photons generated due to graviton interaction with the detectors specifically reads $n_\Gamma=\Gamma_{if}\tau$. The energy density due to $n_\Gamma$ number of photons read $u_\Gamma=\frac{U_\Gamma}{V}=\frac{n_\Gamma\hbar\Omega_P}{V}$. The change in the energy density is related to the generate electric field by the relation $u_\Gamma=\frac{\epsilon_0E_\Gamma^2}{2}$ which is related to the magnetic field by the relation $B_\Gamma=\frac{E_\Gamma}{c}=\sqrt{\frac{2 u_\Gamma}{\epsilon_0c^2}}=\sqrt{\frac{2 n_\Gamma\hbar\Omega_P}{\epsilon_0c^2V}}$. The signal flux is then obtained by 
\begin{equation}\label{SQUID.2.OTM}
\Phi_{\text{Signal}}=B_\Gamma \mathcal{A}_P=\mathcal{A}_P\sqrt{\frac{2 \Gamma_{if}\tau\hbar\Omega_P}{\epsilon_0c^2L^3}}.
\end{equation}
For a reasonable detection scenario $\Phi_{\text{Signal}}\geq\Phi_{\text{Noise}}$, which helps us to get the strain amplitude to be
\begin{equation}\label{SQUID.3.OTM}
\mathcal{f}_{\text{min}}^{\text{SQUID}}=\frac{4\epsilon_0 L^3 c}{\mathcal{A}_P\omega q\tau}\sqrt{\frac{m_0\omega_0}{2\pi\hbar\nu_F\tau(n_{P_i}+1)}}\mu\Phi_0
\end{equation}
where we propose the use of highly sensitive SQUID set-up. Now such highly sensitive SQUID devices have a very small pick up area and it can be significantly increased using superconducting flux focussing models \cite{SuperconductingFlux}. We consider the pickup area to be $\mathcal{A}_{P}=2.25\times 10^{-6}$ $\text{m}^2$ which allows for the strain sensitivity to be $\mathcal{h}_{\text{min}}^{\text{SQUID}}\simeq 1.17\times10^{-27}$ $\text{Hz}^{-\frac{1}{2}}$. This strain corresponding to the SQUID detection scenario is almost four orders of magnitude smaller than our detector strain sensitivity allowing for a reasonable detection scenario for gravitational waves with optical pumping $n_{P_i}\sim 10^{18}$.
It is therefore quite reasonable to understand that our proposed detector model allows for a reasonable strain sensitivity for gravitational wave detection. Finally, we plot our characteristic strain sensitivity curve against the signal frequency and investigate the dependence on optical pumping in Fig.(\ref{Strain_Sensitivity_OTM}).
For the sensitivity plot, we have considered the detector strain and combined it with the strain corresponding to the SQUID, strain due to mechanical thermal fluctuations and photon shot noise. The temperature considered is $T=1nK$ for the plots. The spikes denote the thermal fluctuations from the mechanical oscillator which is plotted using the strain sensitivity formula corresponding to the fluctuation dissipation theorem\footnote{\textcolor{blue}{For a detailed discussion on the thermal noise as well as the shot noise please refer to the supplementary material.}}. The thermal fluctuations can be significantly reduced via considering a multi array and multi-oscillator set-up. We can also observe, that the sensitivity for the detector becomes more competitive while the optical pumping becomes more significant and the sensitivity even for low optical pumping gives a very reasonable sensitivity value near the resonance points, however, to obtain a more rigorous noise profile model one also needs to consider seismic noise as well as cavity decay rate which is beyond the scope of our current manuscript.

\section{Discussion and Conclusion}\label{Conclusion}
In this work, we propose a fundamental and novel proposal for tabletop gravitational wave detection using the interaction of charged harmonic oscillators with photons and classical gravitational wave fluctuations. We start with the relativistic action for a resonant Weber bar in presence of gravitational wave fluctuation and combine it with the Maxwell action in curved background. We then quantize the matter as well as the electromagnetic part of the model system by raising the phase space variables of the harmonic oscillator as well as the electromagnetic field and imposing suitable canonical commutation relation among the conjugate variables. The important thing to remember is that the gravitational wave is treated classically here. We then obtain the analytical form of the Hamiltonian operator which helps us to identify the interaction terms involving the detector-photon, photon-gravitational waves,  gravitational wave-detector, and most importantly the detector-photon-gravitational wave coupling term. While the photon-Gravitational wave coupling term results in the conversion of gravitational wave into two photons, we find out that the three point interaction term of the Hamiltonian actually results in a more physically involved scenario. We find out that the three point interaction Hamiltonian allows the mechanical oscillator to act as a transducer resulting in a conversion of the gravitational wave into a low frequency photon while the oscillator excites and jumps one energy level. From the transition rate, we find out that the spontaneous emission rate is quite low, however, it can be enhanced using an array of identical oscillators. If $\mathcal{N}=100$ oscillator array is created, an overall gain of the order of $10^4$ can be achieved in a realistic experimental scenario. It is still more feasible to look at the stimulated emission scenario while the initial state is pumped using low frequency photons which allows for a measurable transition rate of the order of $0.1-10^2$ $\text{sec}^{-1}$. We have then proposed a novel experimental proposal using a multi-array system of stimulated emitters where each emitter array consists of identical charged harmonic oscillators while the entire system is placed inside of a cavity with the cavity walls being formed of low-frequency harmonic oscillators with the oscillation frequency being equal to the difference of the gravitational wave and the small identical harmonic oscillators. Electromagnetic waves are then pumped in high intensity for allowing stimulated emission of photons from the cavity walls with frequency $\Omega_P=\omega-\omega_0$ which is then stopped before the detector starts interacting with the incoming gravitational wave frequency. After the interaction of the gravitational wave with the detector the detector emits photons which is then converted to direct current using a superconducting quantum interference device or SQUID allowing for the indirect detection of classical gravitational wave signal. We have then provided a strain sensitivity analysis as well as a noise-profile analysis for our proposed experimental detector and find out that the sensitivities become comparable to the interferometer based gravitational wave detectors like LIGO and VIRGO. We finally plot the sensitivity curve against frequency considering all possible noise sources in Fig.(\ref{Strain_Sensitivity_OTM}) and demonstrated the feasibility of using such a novel tabletop experimental set-up as a new gravitational wave detector model.

\onecolumngrid
\section*{Supplementary material}
\begin{adjustwidth}{60pt}{60pt}
Here we provide detailed derivation and added explanations of some of the analytical results presented in the primary manuscript ``Gravity mediated entanglement of phonons in Bose-Einstein condensates".
\end{adjustwidth}
\section{Response function for the mechanical oscillator}
We systematically derive the response function for the mechanical oscillator. The Lagrangian for the model system following the main text file reads
\begin{equation}\label{S.1}
\begin{split}
L=&\frac{m_0}{2}\left(\dot{\xi}^2-\dot{\bar{h}}_{xx}(t,0)\dot{\xi}\xi-\omega_0^2\xi^2\right)+q_PA\dot{\xi}+\frac{m_P}{2}(1-\bar{h}^{xx}(t,0))(\dot{A}^2-\omega_P^2A^2)~.
\end{split}
\end{equation}
The Euler-Lagrange equation for the detector sector is then obtained to be
\begin{equation}\label{S.2}
\begin{split}
\frac{d}{dt}\left(\frac{\partial L}{\partial\dot{\xi}}\right)-\frac{\partial L}{\partial \xi}&=0\\\implies \ddot{\xi}(t)+\omega_0^2\xi(t)+\frac{q_P}{m_0}\dot{A}(t)&=\frac{1}{2}\ddot{\bar{h}}_{xx}(t,0)\xi(t)
\end{split}
\end{equation}
The right-hand side is the gravitational force term and up-to a good approximation can be replaced by $\frac{1}{2}\ddot{\bar{h}}_{xx}(t,0)\xi_0$. We now need to determine $A(t)$ while the time evolution of the vector field can be neglected. From the Euler Lagrange equation $\frac{d}{dt}\left(\frac{\partial L}{\partial\dot{A}}\right)-\frac{\partial L}{\partial A}=0$, we then obtain
\begin{equation}\label{S.3}
\ddot{A}(t)+\omega_P^2A(t)-\dot{\bar{h}}_{xx}(t,0)\dot{A}(t)\simeq \frac{q_P}{m_P}\dot{\xi}(t)+\frac{q_P}{m_P}\bar{h}_{xx}(t,0)\dot{\xi}(t)~.
\end{equation}
Ignoring the Gertsenshtein coupling directly between the photon and gravitational waves and comparing the $\mathcal{O}(h)$ terms, we obtain a relation between $\dot{\xi}(t)$ and $\dot{A}(t)$ as (dropping $\mathcal{O}(h^2)$ contributions)
\begin{equation}\label{S.4}
\dot{A}(t)\simeq-\frac{q_P\bar{h}_{xx}(t,0)}{m_P \dot{\bar{h}}_{xx}(t,0)}\dot{\xi}(t)
\end{equation}
where we have used the template for the gravitational wave $\bar{h}_{xx}(t,0)=2f_0\cos(\omega t)$ from the main manuscript file. Substituting this expression in eq.(\ref{S.2}), defining $\gamma\equiv \frac{q_P^2}{m_0m_P\omega}$, and substituting $\bar{h}_{xx}(t,0)=2f_0\cos(\omega t)$ , we can write down the Euler-Lagrange equation in eq.(\ref{S.2}) as
\begin{equation}\label{S.5}
\ddot{\xi}(t)+\gamma \dot{\xi}(t)+\omega_0^2=-\omega^2\xi_0f_0\cos(\omega t)~.
\end{equation}
It is important to note that the damping factor comes to be time dependent which is an artefact coming due to the division by $h$ and can be completely removed if the equation is expressed in the Fourier space. The above expression exactly represents a damped-driven harmonic oscillator. This lets us solve the equation exactly and, we obtain an analytical expression for $\xi(t)$ as
\begin{equation}\label{S.6}
\begin{split}
\xi(t)=&\frac{f_0\xi_0\omega^2\left(\left(\omega^2-\omega_0^2\right)\cos(\omega t)-\gamma\omega \sin(\omega t)\right)}{\left(\omega^2-\omega_0^2\right)^2+\gamma^2\omega^2}\\
=&\mathcal{A}_{\text{RD}}(\omega)\cos(\omega t)+\mathcal{B}_{\text{RD}}(\omega)\sin(\omega t)
\end{split}
\end{equation}
with $\mathcal{A}_{\text{RD}}$ and $\mathcal{B}_{\text{RD}}$ being defined as
\begin{equation}\label{S.7}
\mathcal{A}_{\text{RD}}(\omega)=\frac{f_0\xi_0\omega^2\left(\omega^2-\omega_0^2\right)}{\left(\omega^2-\omega_0^2\right)^2+\gamma^2\omega^2}~\text{and}~~\mathcal{B}_{\text{RD}}(\omega)=-\frac{\gamma f_0\xi_0\omega^3}{\left(\omega^2-\omega_0^2\right)^2+\gamma^2\omega^2}~.
\end{equation}
The amplitude $|\xi(t)|$ is then obtained to be $|\xi(t)|=\sqrt{\mathcal{A}_{\text{RD}}^2+\mathcal{B}_{\text{RD}}^2}=\frac{f_0\xi_0\omega^2}{\sqrt{\left(\omega^2-\omega_0^2\right)^2+\gamma^2\omega^2}}$ and the corresponding response function reads
\begin{equation}\label{S.8}
\mathcal{R}(\omega)=\frac{|\xi(t)|}{\xi_0}=\frac{f_0\omega^2}{\sqrt{\left(\omega^2-\omega_0^2\right)^2+\gamma^2\omega^2}}~.
\end{equation}
The important thing to understand is that $\gamma=\frac{q_P^2}{\epsilon_0 m_0\omega_0 m_P\omega}=\frac{q^2}{\epsilon_0 m_0\omega_0 L^3 \omega}$ and as a result $\gamma\omega=\frac{q^2}{\epsilon_0 m_0\omega_0 L^3}$ which is independent of the gravitational wave frequency. This helps use redefined a new damping factor $\gamma_0$ such that $\gamma_0=\frac{\gamma\omega}{\omega_0}$ and the quality factor reads
\begin{equation}\label{S.9}
Q_0=\frac{\omega_0}{\gamma_0}=\frac{\omega_0^2}{\gamma\omega}\implies \gamma\omega =\frac{\omega_0^2}{Q_0}.
\end{equation}
We can then rewrite the response function as
\begin{equation}\label{S.10}
\mathcal{R}(\omega)=\frac{|\xi(t)|}{\xi_0}=\frac{f_0\omega^2}{\sqrt{\left(\omega^2-\omega_0^2\right)^2+\frac{\omega_0^4}{Q_0^2}}}
\end{equation}
where the denominator is now in the well known form of damped driven harmonic oscillators. 
\subsection{Thermal strain sensitivity}
One can now read of the modulus square of the mechanical susceptibility (the mechanical susceptibility $\chi(\omega)$ has the dimension $\text{M}^{-1}\text{T}^2$) as
\begin{equation}\label{S.11}
|\chi(\omega)|^2=\frac{1}{m_0^2\left(\left(\omega^2-\omega_0^2\right)^2+\frac{\omega_0^4}{Q_0^2}\right)}~.
\end{equation}
The fluctuation dissipation theorem gives the analytical expression for the Force noise power spectral density as \cite{FD1,FD2}
\begin{equation}\label{S.12}
S_{\text{FD}}(\omega)=2 m_0\gamma\hbar \omega\coth\left(\frac{\hbar\omega}{2k_BT}\right)=\frac{2 m_0\hbar \omega_0^2}{Q_0}\coth\left(\frac{\hbar\omega}{2k_BT}\right)~.
\end{equation}
The thermal sensitivity is then obtained to be
\begin{equation}\label{S.13}
S_{\text{Th}}(\omega)=|\chi(\omega)|^2S_{\text{FD}}(\omega)=\frac{2\hbar\omega_0^2}{m_0Q_0\left(\left(\omega^2-\omega_0^2\right)^2+\frac{\omega_0^4}{Q_0^2}\right)}\coth\left(\frac{\hbar\omega}{2k_B T}\right)
\end{equation}
with $k_B$ denoting the Boltzmann constant and $T$ denoting the temperature of the cryogenic cavity-QED set-up. Then the pure thermal strain sensitivity is obtained by
\begin{equation}\label{S.14}
\mathcal{h}_{\text{Th}}=\sqrt{\frac{S_{\text{Th}}(\omega)}{L^2}}
\end{equation} 
which has the usual dimension of $\text{Hz}^{-\frac{1}{2}}$.
\subsection{Shot noise}
The shot noise follows a Poisson distribution and therefore the change in the counting of photons is effectively $\delta n_{P}^\Gamma=\sqrt{n_{P}^\Gamma}$ which helps us to obtain $\frac{\delta n_{P}^\Gamma}{n_{P}^\Gamma}=\frac{1}{\sqrt{n_P^{\Gamma}}}$. The number of photons emitted in the cavity in time $\tau$ reads $n_{P}^{\Gamma}=\Gamma_{if}\tau=\frac{\omega^2q^2f_0^2\tau^2}{16\epsilon_0L^3\Omega_Pm_0\omega_0}(n_{P_i}+1)$. Now, for minimum strain sensitivity $f_0$ can be replaced using $h$ which lets us obtain
\begin{equation}\label{S2.15}
\begin{split}
\frac{\delta n_P^\Gamma}{n_{P}^\Gamma}=2\frac{\delta h}{h}=&\frac{1}{\sqrt{n_P^\Gamma}}\\
\implies \delta h=&\frac{h}{2\sqrt{n_{P}^\Gamma}}=\frac{2L}{\omega q \tau}\sqrt{\frac{\epsilon_0 Lm_0\omega_0\Omega_P}{n_{P_i}+1}}~.
\end{split}
\end{equation}
The strain sensitivity is then simply obtained by $S_h^{\frac{1}{2}}=\delta h\sqrt{\tau}$ and the shot noise strain sensitivity can be represented in terms of the detector susceptibility as
\begin{equation}\label{S2.16}
\mathcal{h}_\text{Shot}=\frac{\delta h\sqrt{\tau}}{m_0\omega_0^2|\chi(\omega)|}~.
\end{equation}
while $m_0\omega_0^2|\chi(\omega)|$ gives the normalized oscillator response. One can directly use the response $|\mathcal{R}(\omega)|$ but it leads to incorrect noise curves because of the existence of gravitational amplitude.

\end{document}